\documentclass[%
 aip,
 amsmath,amssymb,
 reprint,%
]{revtex4-1}

\usepackage{graphicx}
\usepackage{dcolumn}
\usepackage{bm}

\usepackage[utf8]{inputenc}
\usepackage[T1]{fontenc}
\usepackage{mathptmx}

\usepackage[usenames]{color}
\usepackage{graphicx}
\usepackage[caption=false]{subfig}
\usepackage{wasysym} 
\usepackage{xfrac}
\usepackage{float}
\usepackage{comment}

\begin{document}

\preprint{AIP/123-QED}

\title[Microstructural control of the transport properties of $\beta$-FeSe films grown by sputtering]{Microstructural control of the transport properties of $\beta$-FeSe films grown by sputtering}

\author{M. V. Ale Crivillero}
\altaffiliation[Present Address]{: Max Planck Institute for Chemical Physics of Solids, Dresden, Germany}
\email{victoria.ale@ib.edu.ar.}
\author{M. L. Amig\'o}%
\altaffiliation[Present Address ]{: Institute for Solid State Research, IFW Dresden, Germany.}
\author{N. Haberkorn}%
\author{G. Nieva}%
\author{J. Guimpel}%

\affiliation{ 
Divisi\'on Bajas Temperaturas, Centro At\'omico Bariloche, CNEA - CONICET - Instituto Balseiro, UNCuyo y CNEA, (8400) Bariloche, Argentina.
}%

\date{\today}

\begin{abstract}
We have investigated the correlation between structural and transport properties in sputtered $\beta$-FeSe films grown onto SrTiO$_3$ (100).
The growth parameters, such as substrate temperature and thickness, have been varied in order to explore different regimes.
In the limit of textured thick films, we found promising features like an enhanced $T_{\rm c}\sim12\,$K, a relatively high $H_{\rm c2}$ and a low anisotropy. By performing magnetoresistance and Hall coefficient measurements, we investigate the influence of the disorder associated with the textured morphology on some features
attributed to subtle details of the multi-band electronic structure of $\beta$-FeSe.
Regarding the superconductor-insulator transition (SIT) induced by reducing the thickness, we found a non-trivial evolution of the structural properties and morphology associated with a strained initial growth and the coalescence of grains.
Finally, we discuss the origin of the insulating behavior in high-quality stressed epitaxial thin films. We found that a lattice distortion, described by the Poisson's coefficient associated with the lattice parameters \textit{a} and \textit{c}, may play a key role.

\keywords{Thin Films, $\beta$-FeSe, Sputtering}

\end{abstract}

\maketitle

\section{\label{intro}Introduction}

Among the iron-based superconductors, iron selenide stands out as a rich system for studying new emergent phenomena. Particular interest in this material has been triggered by the strong enhancement of the superconducting critical temperature with pressure (hydrostatic \cite{Sun2016,Medvedev2009} or residual stress \cite{Nabeshima2013,Imai2016}) and the reports of the opening of a superconducting gap above $65\,$K in extremely thin FeSe films grown by MBE \cite{Liu2012,Wang2012}. In the last case, interface charging and/or electron-phonon coupling are proposed as the mechanism responsible for raising the critical temperature. From the fundamental point of view, interesting behavior has been reported in this material such as a multi-band electronic structure \cite{Lei2012,Watson2015}, tetragonal to orthorhombic \cite{McQueen2009} and nematic phase transitions at $90\,$K \cite{BohmerPRB13,BohmerPRL15}. 

It is also a promising candidate for applications given its high-field performance and as a candidate for the construction of cryogenic sensors. The
potential use of Fe-based superconductors in the electronics sector has been discussed by Haindl et al \cite{Haindl2014}. A particular perspective niche is the use of FeSe in the construction of Microwave Kinetic Inductance Detectors (MKID's) \cite{Mazin}. The relatively high critical temperature of $9\,$K would allow its use at temperatures in the range of $^4$He instead of miliKelvins, as occurs with the currently used materials. However, the integration in these devices requires the growth of high-quality thin films. 
Given this, it is important to
understand the growth mechanism and how it affects the physical properties and morphology of the resulting material.

Despite the structural simplicity of this compound, high-quality bulk single crystals were obtained only years after its discovery \cite{Hsu2008,Chareev}.  The main feature of the temperature dependence of the resistivity is a superconducting transition at $9\,$K followed by a semimetallic state at higher temperatures.
Motivated for the potential technical applications \cite{Haindl2014}, the challenge of growing thin films has been also faced up. Thin films of FeSe have been successfully synthesized by different growth methods such as pulsed laser deposition (PLD) \cite{Nabeshima2013,Imai2016,Nie2009,Ta-Kun-Chen2009}, molecular beam epitaxy (MBE) \cite{Sun2016,Liu2012,Tan2013,Song2011} and in a lesser extent by sputtering \cite{SchneiderPRL,SchneiderSUST,SpellerSUST,Venzmer2016}. Nevertheless, there are discrepancies between different authors with the transport properties of epitaxial films in the limit of small thicknesses, irrespective of the growth method. There are reports of both superconducting and semiconducting/insulating behavior.
Since controlling the stoichiometry and/or the crystalline structure is still a challenging issue, a remaining question is whether minor structural or compositional changes are relevant.
In the case of films grown by sputtering, this implies a superconductor-insulator transition (SIT) as a function of thickness \cite{SchneiderPRL,Wang2DMater}. This kind of behavior has been discussed within a disordered granular superconductor scenario \cite{SchneiderPRL} where the disorder associated with smaller thickness induces a superconductor-insulator transition. In this context, fundamental questions concerning the mechanism governing the transition are still open.

In this work, we report on the fabrication of thin films of $\beta$-FeSe by dc magnetron sputtering from stoichiometric targets, with the aim to study the influence on the transport properties of the structural characteristics in the atomic and mesoscopic scales. 
The phase purity and crystal structure of the films were characterized by X-ray diffraction measurement (XRD). The surface morphology was studied using a Scanning Electron Microscope (SEM) and an Atomic Force Microscope (AFM). The local chemical composition was measured by Energy-Dispersive X-ray Spectroscopy (EDX) and Rutherford Backscattering Spectrometry (RBS). The magnetization was measured in a commercial SQUID magnetometer.
We carried out electrical transport measurements with magnetic fields up to $16\,$T.

\section{Experimental details}

Fe$_{1-x}$Se thin films have been deposited by dc magnetron sputtering. 
The polycrystalline target was fabricated from Fe ($99.95\,$\%) and Se ($99.999\,$\%) grits. Two consecutive cycles of mixing, heating up to $430\,^{\circ}$C for $48$ hours
in an evacuated quartz tube,
and milling were performed to favor the solid state reaction of the powder.
Low heating rates were used while the temperature was close to the melting point of Se.
In a final process, the powder was pressed and sintered at $430\,^{\circ}$C for 4 days. X-ray diffraction and Rietveld refinement were performed at each stage to characterize the powder.
The final target material was $\beta$-FeSe with approximately $1.5\,$\% $\gamma$-Fe$_7$Se$_8$. A very small amount, \textless5\,\permil, of FeO was also observed.

Films were grown onto SrTiO$_3$ (100), labelled from now on as STO. The substrates were attached to a resistive heater with Ag paste, which allowed for heating up to a nominal temperature of 680$^{\circ}$C \cite{heater}. The substrate temperature, $T_{\rm s}$, was measured and controlled with a thermocouple in direct thermal contact with the substrates. The base pressure of the sputtering system was $3 \times 10^{-6}$\,Torr. The growth parameters that were kept constant for all films are sputtering gas, 99.999\% pure Ar, and pressure, $45$\,mTorr, target power, $8$\,W for a $1.5$\,"  diameter target, and target to substrate distance, at $5$\,cm. In this conditions, the growth rate is approximately $13.5\,$nm/min. These parameters were found to be optimal in previous, unreported tests. 

\begin{figure}[!t]
   \centering
       {\includegraphics[width=0.48\textwidth]{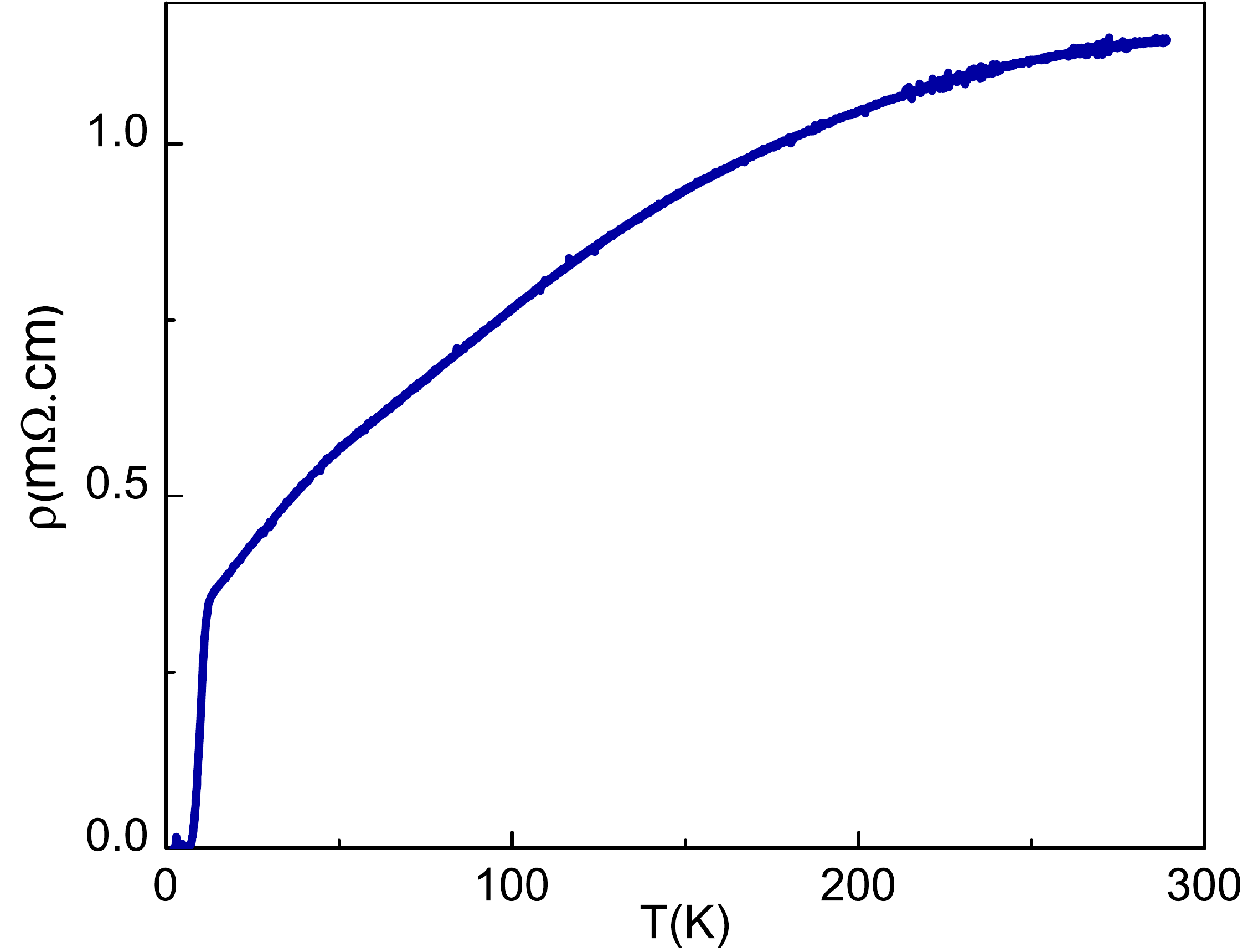}
     }
     
     \caption{In plane resistivity, $\rho$, versus temperature, T, for a superconducting $400\,nm$ thick $\beta$-FeSe film grown onto STO.}
     \label{FigTransporte}
   \end{figure}

The chemical composition of the films has been routinely measured by EDX and for some samples also measured by Rutherford Backscattering Spectrometry (RBS). Under the conditions described above, all the samples have a composition close to the expected values, i.e. Fe:Se$\,\approx\,$1:1, within the error in EDX analysis. The homogeneity of the films was also investigated by performing several spot and area analysis over different regions. No spatial variations were found.

The thickness of the films was measured by contact profilometry in well-defined steps. These steps were generated by first masking part of the substrates with silver epoxy, growing the films, and finally removing the epoxy. For thin films, the thickness was also determined through low angle X-Ray reflectivity, XRR, measuring the periodicity of the Kiessig fringes \cite{Kiessig1931}. When the later were observed, both methods of thickness determination coincided within 10 \%.

\begin{figure}[t!]
 \centering
 \includegraphics[width=0.48\textwidth]{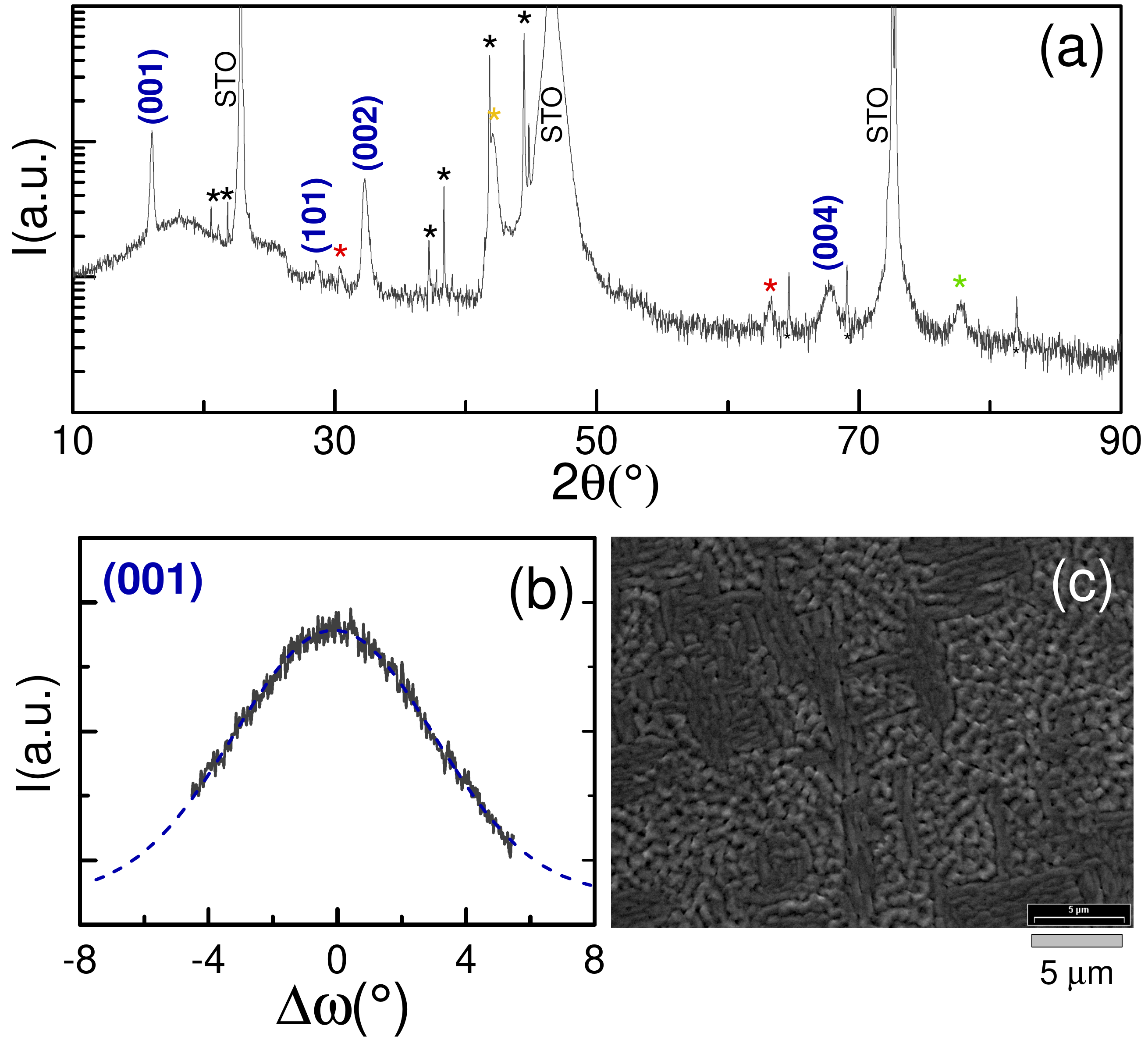}
 \caption{Structural and morphological characterization of a $400\,$nm $\beta$-FeSe grown onto STO.
 (a) $\theta-2\theta$ XRD pattern: The (00l) peaks associated with the $\beta$-FeSe phase are clearly observed. The red labels correspond to (00l) $\gamma$-Fe$_7$Se$_8$. The yellow label could be either indexed as (102) $\gamma$-Fe$_7$Se$_8$ or (200) FeO, while the green label might be identify as (203) $\gamma$-Fe$_7$Se$_8$. The black asterisk signs stand for spurious peaks related to the substrate; i.e. peaks measured in a pristine substrate which are no expected in an ideal single crystal.  (b) Rocking curve of the (001) $\beta$ phase peak.
 (c) SEM image of the topography of the $400\,$nm $\beta$-FeSe.}
 \label{FigRX}
\end{figure}

The diffraction patterns were measured at room temperature with two different diffractometers: a PANalytical Empyrean diffractometer or a Phillips PW 3710
in Bragg-Brentano geometry with Cu K$_{\alpha}$ radiation.
In the former case, the samples were mounted in a 4-circle Eulerian cradle, which allowed for precise alignment. Due to physical limitations of the Eulerian cradle, a monochromator could not be used. Instead, a Ni filter was used to diminish the Cu K$_\beta$ radiation. For the extremely intense substrate diffraction, the K$_\beta$ peak and the Ni absorption edge are visible in the data, the later as a step-like discontinuity in the background intensity between the K$_\alpha$ and K$_\beta$ peaks.

Electrical transport properties were measured in a standard 4 probe geometry, defined by UV photo-litho\-graphy and ion milling. For the Hall effect measurements, two probes at both sides of the current carrying line were added. The measurements were performed in an Oxford cryostat equipped with an $18$\,T superconducting coil and sample rotation capability, in the $1.8$\,K to $300$\,K range. 

\section{Results and discussion}

Since previous systematic studies showed a complex influence of growth conditions on the structural and physical properties of the films, we choose the growth parameters, such as substrate temperature and thickness, in order to explore different regimes.
First, we focus on textured thick films with optimal superconducting properties.
Second, regarding the superconductor-insulator transition (SIT) induced by reducing the thickness, we performed a comprehensive study of the evolution of the structural properties and morphology.
Finally, we discuss the origin of the insulating behavior on epitaxial stressed films.

\subsection{\label{Superconducting properties}Superconducting thick films}

Since the properties which are sought to optimize for integrating FeSe films in devices are the superconducting ones, first we will focus on what we found to be the ``best" superconducting properties. The optimal parameters for growing a superconducting film on STO, i.e. those which produce the highest, $T_{\rm c}^{\rm onset}\simeq12\,K$, and sharpest, $\Delta T_{\rm c}\simeq4\,$K, superconducting transition are $T_{\rm s}=530^\circ$C and thickness $t\sim400\,$nm.
These values of superconducting onset are slightly higher than the previously reported for
macroscopic high-quality samples, $\sim9\,$K \cite{Chareev}, and
sputtered films, $\sim10\,$K \cite{SchneiderPRL,SchneiderSUST,Venzmer2016}. 
As a representative example of the temperature dependence of the electrical resistivity,
Figure \ref{FigTransporte} shows $\rho(T)$ in the $2-300\,$K range of one of these films.
Besides the superconducting transition at low temperature, a semimetallic like behavior is observed in the normal state
with a negative curvature up to room temperature.

Regarding the crystalline structure, Figure \ref{FigRX}(a) shows the X-ray diffraction data for the same film. The most intense peaks can be indexed as the (00l) $\beta$-FeSe family. The rest of the peaks can be identified as substrate related or low intensity (101) $\beta$-FeSe and (001) $\gamma$-Fe$_7$Se$_8$. The (001) peak rocking curve has a full width at half maximum (FWHM) of approximately 5$\,^\circ$, indicating textured growth (Figure \ref{FigRX}(b)). These results indicate preferred c-axis growth, with coexisting small inclusions of (101) $\beta$-FeSe and (001) $\gamma$-Fe$_7$Se$_8$.
The SEM topography, Figure \ref{FigRX}(c), shows a tweed grain pattern with a characteristic length around $1\mu$m.
We previously reported a similar increase of $T_{\rm c}$ for bulk crystals with a nanoscale intergrowth of $\beta$-FeSe and $\gamma$-Fe$_7$Se$_8$ \cite{Amigo2017}.
Since there are also inclusions in the sputtered films, the enhancement could be related to the presence of this type of defect. 
However, another possibility which cannot be neglected is the existence of local tensile stress due to grain matching.

\begin{figure}[!t]
   \centering
       {\includegraphics[width=0.48\textwidth]{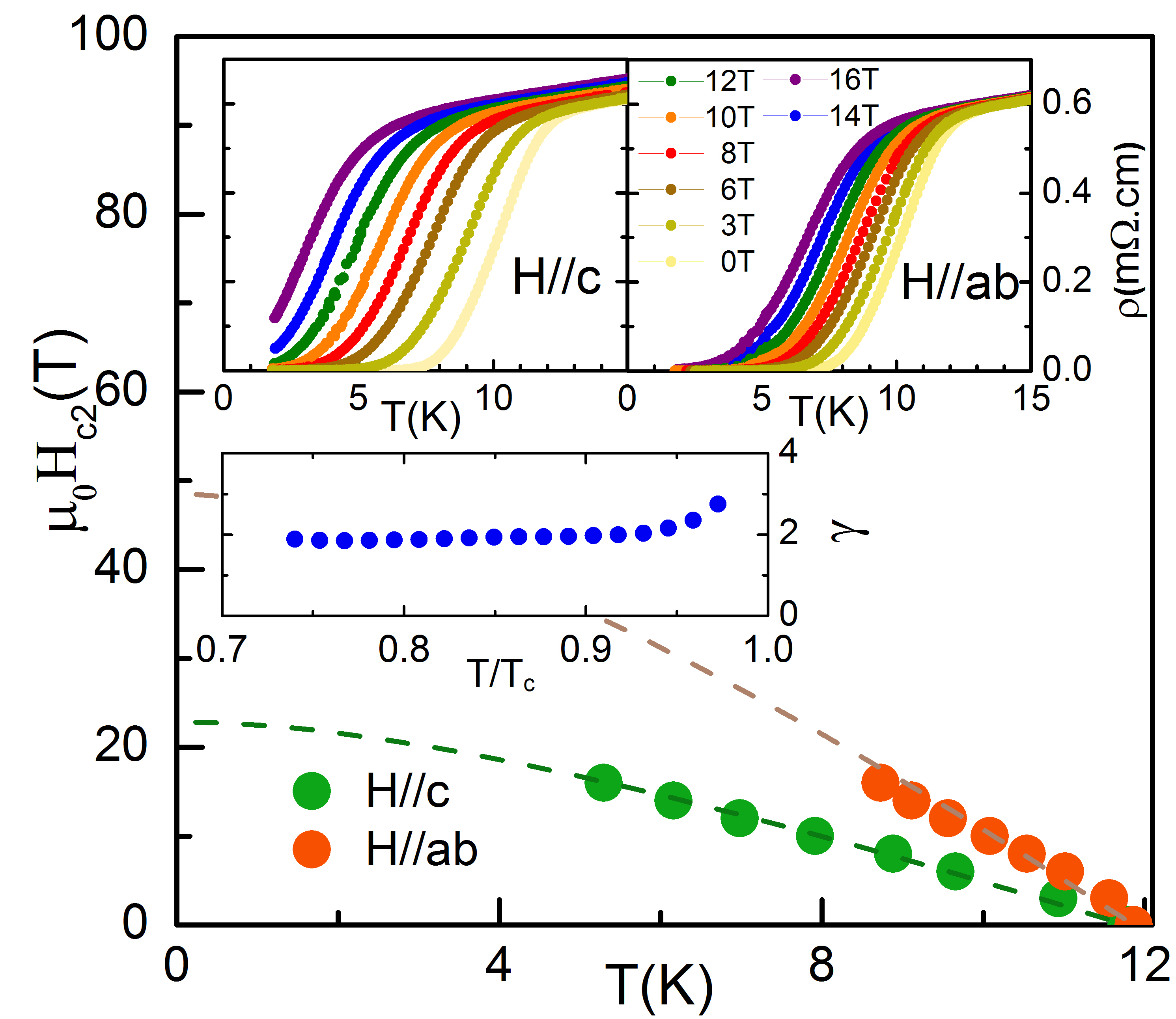}
     }
          \caption{Superconducting phase diagrams, $\mu_0H_{\rm c2\perp}(T)$ and $\mu_0H_{\rm c2\|}(T)$, for a $400\,$nm $\beta$-FeSe films grown onto STO. A magnetic field up to $16\,$T was applied parallel and perpendicular to the substrate surface. The dotted lines correspond to the dependence predicted by the WHH model. Top insets show the superconducting transitions with $H$ as a parameter, while bottom inset displays the temperature dependence of the anisotropy.}
     \label{FigHC2}
   \end{figure}

From the point of view of high-field applications, properties as the critical field and its anisotropy are relevant. The top insets in Figure \ref{FigHC2} show the evolution of the superconducting transition with applied field $H$ parallel and perpendicular to the substrate surface, which corresponds mainly to magnetic field along the \textit{ab} plane and along \textit{c} axis, respectively. From these curves the transition temperatures are determined and so the perpendicular and parallel critical fields, $\mu_0 H_{c2\perp}(T)$ and $\mu_0 H_{c2\|}(T)$, which are shown in the main figure.
The dotted lines correspond to the dependence predicted by the Werthamer-Helfand-Hohenberg (WHH) model.
Determining the slope of $\mu_0 H_{\rm c2}(T)$ in the vicinity of $T_{\rm c}$ and using the WHH model \cite{WHH}, the values of $\mu_0 H_{\rm c2}(T=0)$ are estimated as $\mu_0 H_{\rm c2\perp}(0)=(22.9\pm 0.5)\,$T and $\mu_0 H_{\rm c2\|}(0)=(48.5\pm 0.5\,)$T, in accordance with the previously reported values \cite{Nabeshima2013,Terashima2014}. From the Ginzburg Landau relations for $3$D superconductor in the clean limit \cite{Haindl2014}, the estimated values for coherence lengths at zero temperature are $\xi_{ab}(0)\sim3.8\,$nm and $\xi_{c}(0)\sim1.8\,$nm. Since these values are considerably smaller than the film thickness and the grain size, our use of the 3D Ginzburg Landau equations is justified. Consequently, the difference between $\mu_0 H_{\rm c2\perp}(T)$ and $\mu_0 H_{\rm c2\|}(T)$ can be interpreted as due to the intrinsic $\beta$-FeSe anisotropy. The anisotropy, defined as $\gamma=H_{\rm c2\|}(T)/H_{\rm c2\perp}(T)$, takes a maximum value of $2.4$ at $11.6\,$K and decreases for lower temperatures (see bottom inset). These values are probably underestimated due to the presence of misaligned grains, but they are not far from the reported values. Indeed, this anisotropy lies within the reported curves for pure $\beta$-FeSe bulk crystals and for crystals with impurity phases \cite{Amigo2015}.

\begin{figure}[!t]
   \centering
       {\includegraphics[width=0.48\textwidth]{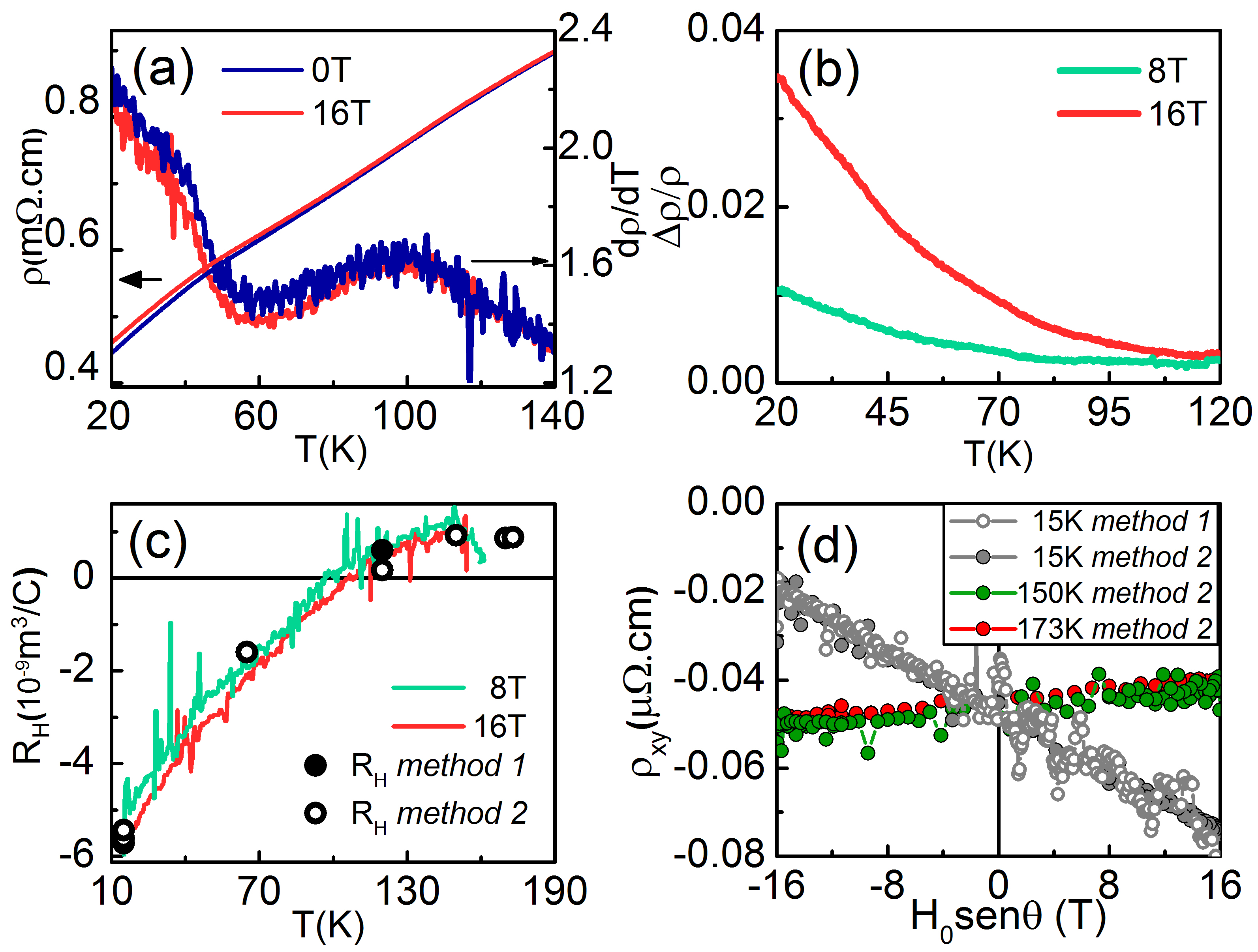}
     }
          \caption{Transport properties in the normal state of a $400\,$nm $\beta$-FeSe films grown onto STO. \textit{(a)} Detail of the in plane resistivity, $\rho$, versus temperature, $T$, in the structural transition region including the derivative $\sfrac {d\rho} {dT}$ and data with an applied magnetic field of $16\,$T perpendicular  to  the  substrate surface. \textit{(b)} Transversal magnetoresistance, $\frac{\Delta\rho}{\rho_0}=\sfrac{(\rho(H)-\rho_0)}{\rho_0}$, as a function of temperature with $\mu_0H=8$ and $16\,$T.
          \textit{(c)} Temperature dependence of the Hall coefficient, $R_H$. The data has been obtained with three different protocols: as $\sfrac{\rho_{xy}(T)}{H}$ at constant $H$ (continuous line), as $\sfrac{d\rho_{xy}}{dH}$ at constant $T$ (method 1) and as $\sfrac{d\rho_{xy}}{dH_{\perp}}$ in an angular dependence experiment at constant $H$ and $T$ (method 2). 
     In all cases, the field is perpendicular to the film surface, except in the case of the angular dependence experiments where $H_{\perp}$ corresponds to the component of the applied field normal to the film.
     \textit{(d)} Field dependence of the Hall resistivity, $\rho_{xy}$, at $T=15$, $150$ and $173\,$K. For $T=15$\,K, the data obtained with methods 1 and 2 are presented.}
     \label{JAP}
   \end{figure}

Coming back to the transport properties in the normal state, our sample mimics some features reported in the literature. The data in Figure \ref{FigTransporte} shows a smooth slope change that begins at $100\,$K and extends to lower temperatures, instead of the usually abrupt slope change at $T\sim90\,$K correlated to the structural transition for $\beta$-FeSe \cite{McQueen2009}.
This feature is zoomed in the Figure \ref{JAP} \textit{(a)}, together with $d\rho/dT$. 
Data with an applied magnetic field of $16\,$T perpendicular to the substrate surface is also included. 
Figure \ref{JAP} \textit{(b)} shows the temperature dependence of the transversal magnetoresistance, $\frac{\Delta\rho}{\rho_0}=\sfrac{(\rho(H)-\rho_0)}{\rho_0}$, with $\mu_0H=8$ and $16\,$T. A positive magnetoresistance is observed below the structural transition,
which increases as the temperature decreases.
We emphasize that this was mainly reported for bulk high-quality single crystals \cite{Kasahara2014,AmigoLT}, although it has also been reported for $160\,$nm thick PLD grown films \cite{arXiv2018}.

More information on the electronic structure is obtained from Hall-effect measurements. The Hall coefficient $R_{\rm H}$, which has been measured with three different protocols (see caption of Fig. \ref{JAP} \textit{(c)}), takes small positive values above the structural transition, and shows a sign reversal to negative values for $T<T^*$, with $T^*\sim90\,$K. This change of sign is reminiscent of the one presented by single crystals close to this transition, which has been associated with the structural distortion and originating in the multiband character of $\beta$-FeSe \cite{Watson2015,Kasahara2014}.
Nevertheless, there is a striking difference at low temperatures; 
below $T^*$
our films present a linear Hall resistivity up to $16\,$T, in contrast to the nonlinear behavior observed in high-quality single crystals \cite{Watson2015}.
Figure \ref{JAP} \textit{(d)} shows the linear dependence of the Hall resistivity, $\rho_{xy}$, with the magnitude of the applied magnetic field at $T=15$, $150$ and $173\,$K.
It is known that $\rho_{xy}(H)$ depends strongly on the nature of the FeSe samples. For instance, bulk single-crystal exhibit non-linear behavior reveling a multiband feature, while exfoliated single-crystalline flakes and thin films show linear behavior \cite{PRL116,arXiv2018}.
We emphasize that some features like the sign reversal of the Hall coefficient and a concomitant positive magnetoresistance are robust in our samples
against the presence of impurity phases and the disorder associated with the grains texture. Overall, the differences found may shed light on what aspects of bulk single crystals are more sensitive to this kind of imperfections. 

In summary, from the point of view of potential applications, the enhanced $T_{\rm c}$, the relatively high $H_{\rm c2}$ and the low anisotropy are promising features of the films grown by sputtering at $T_{\rm s}= 530\,^\circ$C with a thickness around $400\,$nm.
We investigate the influence of the disorder associated with a textured morphology on some transport properties attributed to subtle details of the multi-band electronic structure of $\beta$-FeSe.
In the normal state, we find that some distinctive characteristics of the nematic phase like changes in the resistivity, a positive transverse magnetoresistance and a sign reversal of the Hall coefficient, are robust against the type and degree of disorder present in the films.
Nevertheless, we do not observe other features like a non-linear Hall effect. 

\subsection{\label{SIT}
Superconductor-insulator transition (SIT): Evolution of the crystalline structure and surface morphology}

   \begin{figure}[!t]
   \centering
       {\includegraphics[width=0.48\textwidth]{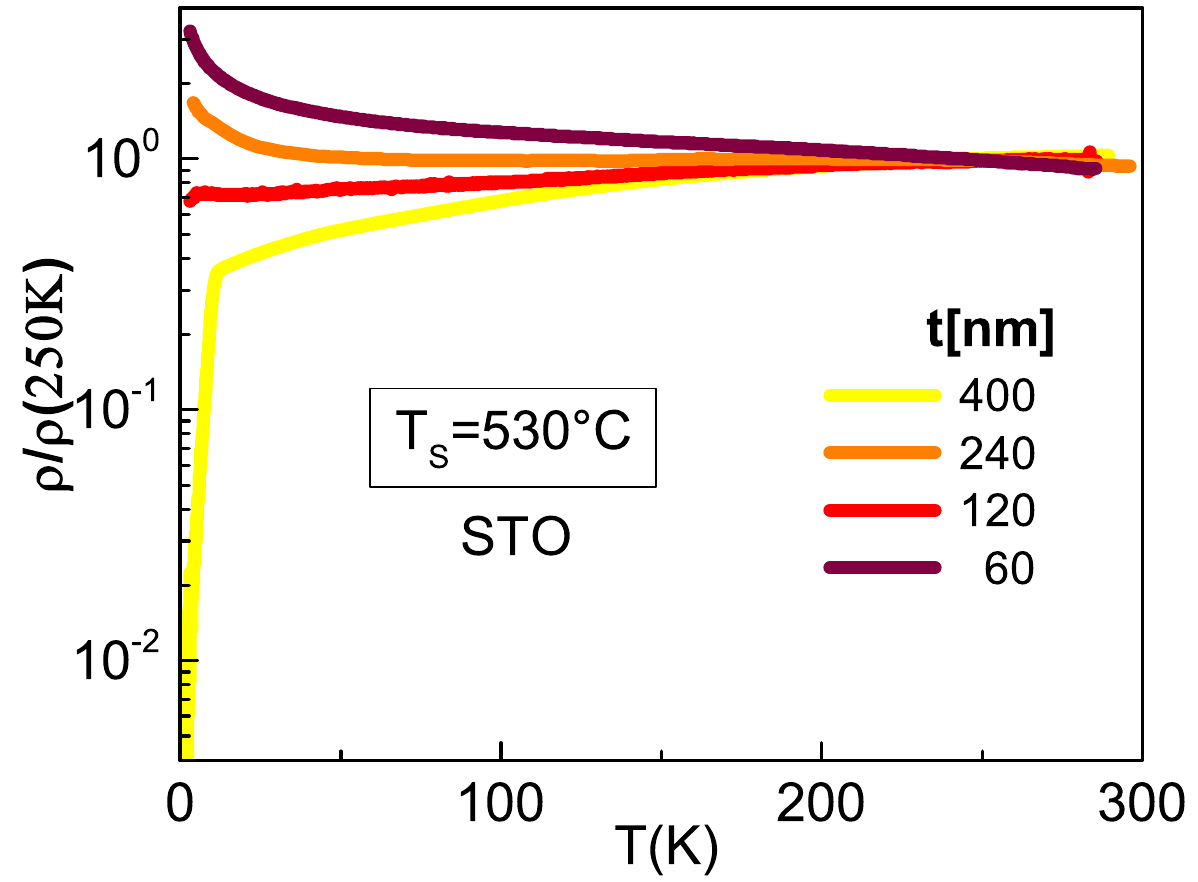}
     }
\caption{Normalized resistivity, $\rho(T)/\rho(250\,\textup{K})$, as function of temperature with thickness as a parameter.
Films were grown at the same $T_{\rm s}=530\,^\circ$C, under equivalent conditions on STO substrates. The thicknesses vary from $60$ to $400\,$nm for growth times between $5$ and $30\,$min.
     }
     \label{FigRvst}
   \end{figure}

What is the effect of thickness on the physical properties of the films? Figure \ref{FigRvst} shows the normalized electrical resistivity, $\rho(T)/\rho(250\,\textup{K})$, as a function of temperature, $T$, for films of different thicknesses. Films were grown at the same $T_{\rm s}=530\,^\circ$C,
under equivalent conditions on STO substrates.
The thicknesses vary from $60$ to $400\,$nm for deposition times between $5$ and $30\,$min.
A change from insulating to superconducting behavior is observed as the film thickness increases,
indicating a superconductor-insulator transition (SIT) \cite{SchneiderPRL,Wang2DMater}.
It is worth to mention that this phenomenology seems to be a general feature of sputtering grown $\beta$-FeSe films. In unreported tests, we observe this crossover
in a wide range of $T_{\rm s}$ ($360\,^\circ$C$\leq T_{\rm s}\leq 630\,^\circ$C), independently of the substrate used (STO or MgO). We found that thin films ($t\leq60\,$nm) present an insulating-like behavior, while the corresponding thicker films ($t\leq300\,$nm) grown under equivalent condition show a metallic-like behavior and a superconductor transition at low temperatures.

\begin{figure}[t!]
 \centering
 \includegraphics[width=0.48\textwidth]{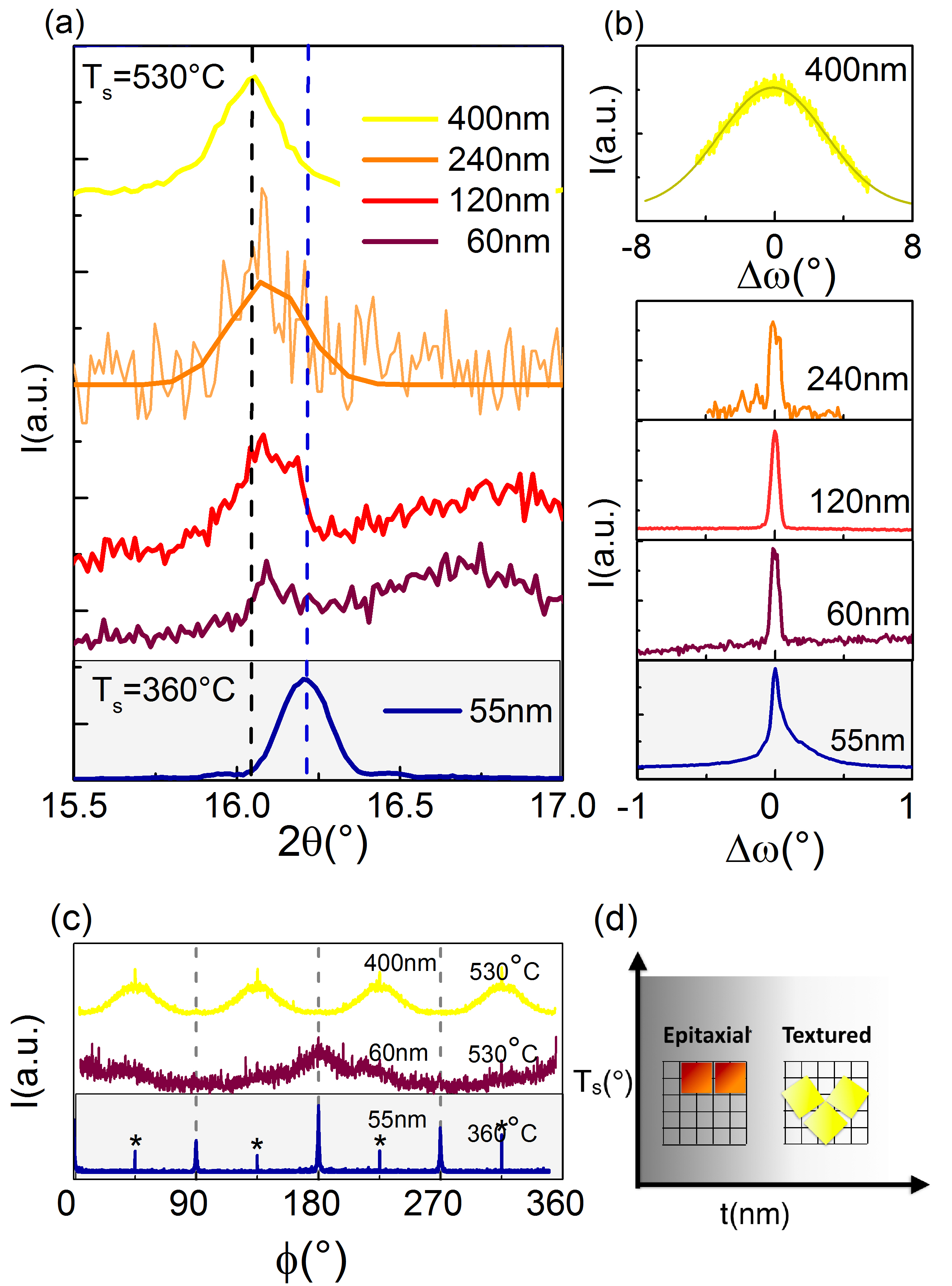}
 \caption{In-plane and out-plane structural evolution in the superconducting-insulating transition. The data correspond to films growth at $T_{\rm s}=530\,^\circ$C.
 Also, the results of an epitaxial thin film ($t=55\,$nm) grown at $360\,^\circ$C has been included for comparison.
 (a) Detail of the $\theta-2\theta$ scans in the $\left[15.5:17.0\right]^{\circ}$ range, with the nominal thickness as a parameter.
 The dotted lines indicate the position of the peaks corresponding to the relaxed (black) and a distorted (blue) structure.
 (b) Rocking curves of the corresponding (001) peaks.
 (c) Azimuthal $\phi$ scans for the (101) $\beta$-FeSe diffraction peak. The vertical dashed lines show the position for the [100] axis of the STO and the asterisks correspond to spurious peaks associated with the substrate. 
 (d) Schematic representation of the evolution of the in-plane texture as a function of thickness.}
 \label{FigInPlaneXRD}
\end{figure}

In order to correlate this behaviour with the structural evolution of the films we also studied the crystalline structure and morphology as a function of thickness. Figure \ref{FigInPlaneXRD}(a) shows a zoomed view of the $\theta-2\theta$ XRD patterns
in the $\left[15.5:17.0\right]^{\circ}$ range for the previously mentioned films. The diffraction pattern from an epitaxial thin film ($t=55\,$nm) grown at $360\,^\circ$C,
which will be studied in detail in the next section,
has been included for comparison. In all the cases, the (001) peak is observed. Only for the case of the epitaxial sample grown at $360\,^\circ$C the $c$ lattice parameter is clearly contracted. On the thick film limit, $t=400\,$nm, the peak has the same position as in the bulk material.
The rest of the samples have almost the same lattice parameter with a small but noticeable tendency to be smaller.
This implies that for higher thicknesses the tensile stress is relaxed.

 The full width at half maximum (FWHM) of the rocking curves (Figure \ref{FigInPlaneXRD}(b)) indicates that the out-plane dispersion increase with the nominal thickness. There is also an evolution of the in-plane structure of the films. Figure \ref{FigInPlaneXRD}(c) shows the azimuthal $\phi$ scans for the (101) $\beta$-FeSe diffraction peak of the films. Vertical dashed lines show the position of the measured peaks in an equivalent azimuthal scan for the (101) diffraction of the STO substrate. The figure includes data for films grown at $T_{\rm s}=530\,^\circ$C with thicknesses  $t=60$\,nm and $t=400$\,nm and a film grown  at $T_{\rm s} = 360\,^\circ$C with $t= 55$\,nm. It is clear that the thinner films present an in-plane alignment with the [100] $\beta$-FeSe axis parallel to the [100] STO axis, which is optimal for $T_{\rm s} = 360\,^\circ$C. However, the thicker films show a rotation in the structure with the [110] $\beta$-FeSe axis parallel to the [100] STO axis. 
 Different non cubic-on-cubic alignments
 have been reported previously \cite{PRL103,APL108}.
 The Figure \ref{FigInPlaneXRD}(d) is a schematic representation of the in-plane evolution as a function of thickness.
 Also, the width of the diffraction peaks is greatly increased after the rotation takes place.
Our results are compatible with an initially stressed cube-on-cube growth
which, after a critical thickness, relaxes through a 45$\,^\circ$ rotation around the vertical axis. This rotation is characterized by an increment of the in-plane and out-plane dispersion.

Regarding the morphology,
in contrast to the tweed grain pattern of the superconducting sample, thinner films
present a smoother surface. Notably, in the case of the sample grown at $T_{\rm s}=360\,^\circ$C
the microstructural study of the surface morphology by SEM and AFM measurements revealed a continuous insulating matrix with small embedded particles ($d\sim100\,$nm).

\begin{figure}[t!]
 \centering
 \includegraphics[width=0.48\textwidth]{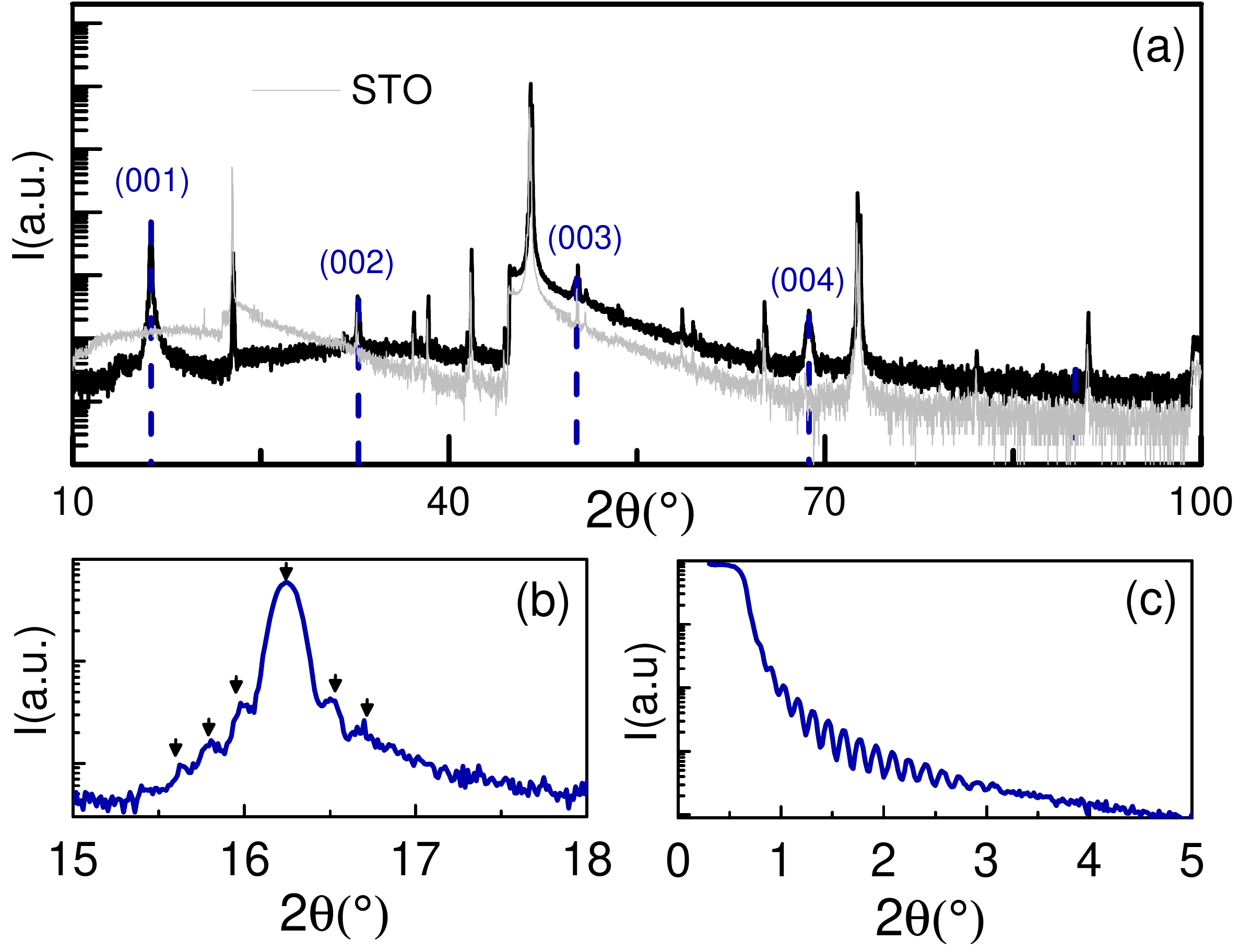}
 \caption{(a) $\theta-2\theta$ diffraction pattern of an epitaxial thin film ($t=55\,$nm) grown at $360\,^\circ$C. For comparison, the corresponding result of a pristine STO substrate is also included.
(b) Detail of the $\theta-2\theta$ scan in the range of the (001) peak. The arrows indicate extra peaks associated with the finite size effect.
(c) Low-angle X-ray reflectivity for the epitaxial thin film.}
 \label{XRD-FeSe050b.pdf}
\end{figure}

In conclusion, concomitant to the superconductor to insulator crossover there is a complex structural and morphological evolution. 
This phenomenology is in contrast to the strong $c$-axis texture irrespective of the film thickness reported by Schneider et al \cite{SchneiderPRL}.
A plausible scenario is a Volmer-Weber island growth mode where the initial layers are characterized by a structural distortion.
By increasing the nominal thickness, the tension relaxes and a regime of oriented grains emerge.
This implies that percolation is a necessary condition for the macroscopic conduction mechanism to reflect the semimetallic and superconducting nature of $\beta$-FeSe. 
A model to understand this phenomenology must be able to explain which is the origin of the insulating behavior in the epitaxial thin film grown at $T_{\rm s}=360\,^\circ$C, despite the lack of islands or cracks and the high-quality structural order.

\subsection{Origin of the insulating behavior}

What could be the origin of the insulating behavior of stressed epitaxial films? To analyze this question, we focus on the epitaxial $\beta$-FeSe ($t=55\,$nm) films grown at $360\,^\circ$C. The high-quality epitaxy of these films is evidenced by:
i) the existence of only (00l) peaks in the diffraction curve (Fig. \ref{XRD-FeSe050b.pdf}a); ii) the sharpness of the out-of-plane rocking curve (Fig. \ref{FigInPlaneXRD}b) and iii) the highly oriented sharp peaks in the $\phi$-scan (Fig. \ref{FigInPlaneXRD}c). Also, a very homogeneous thickness is evidenced by the finite size effect peaks around the (001) peak (Fig. \ref{XRD-FeSe050b.pdf}b) and the presence of Kiessig fringes in the low angle XRR (Fig. \ref{XRD-FeSe050b.pdf}c). From these curves the lattice parameters of these single-phase samples are obtained, showing enlarged $a=3.788\,\textup{\r{A}}$ and contracted $c=5.463\,\textup{\r{A}}$ lattice parameters. This implies a distortion of $\sfrac{\Delta a}{a}=+0.64\,\%$ and $\sfrac{\Delta c}{c}=-1.03\,\%$, with reference to a bulk single crystal ($a=3.771\,\textup{\r{A}}$ and $c=5.521\,\textup{\r{A}}$) \cite{Amigo-Irrad}.

\begin{figure}[t!]
 \centering
 \includegraphics[width=0.48\textwidth]{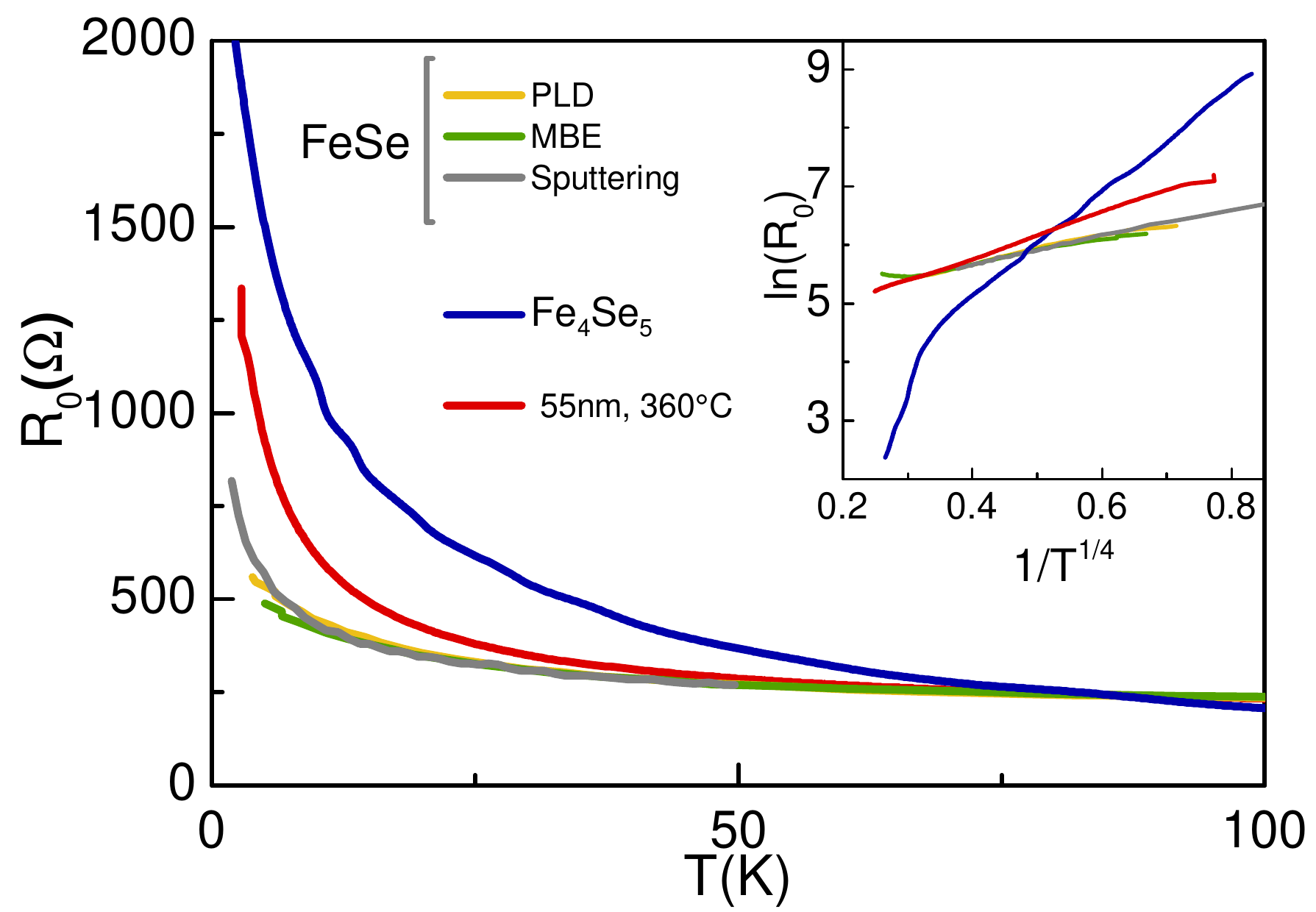}
 \caption{Comparison of the temperature dependence of the normalized electrical resistance of insulating FeSe samples: $R_0$ vs $T$ (main figure) and $\ln R_0$ vs $T^{-\frac{1}{4}}$ (inset). Our data are presented together with those reported in the literature for $\beta$-FeSe samples obtained with different growth methods. The data, normalized to the value of our film at $ 85 \, $ K, were obtained from the digitization of the curves reported by Schneider et al. (Sputtering, \cite{SchneiderPRL}), Hanzawa et al. (MBE, \cite{HanzawaPNAS}), Molatta et al. (PLD, \cite{Molatta2015}) and Chen et al. (nanometric samples of Fe$_{4}$Se$_{5}$ obtained due the extraction of K from K$_2$Fe$_4$Se$_5$ single crystal, \cite{Chen2014}.}
 \label{Transporte_FeSe_semi}
\end{figure}

Since the sample structure is tetragonal, we explore the possibility that the origin of the insulating behavior is due to small structural and/or compositional changes. The first hypothesis considers stoichiometric modifications consistent with the observed crystalline order.
Recently, semiconducting Fe$_{1-x}$Se samples with ordered Fe vacancies have been reported. For $x=0.2$ (Fe$_4$Se$_5$), the phase is still tetragonal ($a=3.76\,\textup{\r{A}}$ and $c=5.47\,\textup{\r{A}}$) but it is not superconducting. It also presents a long-range magnetic order and has been proposed as the non-superconducting parent of FeSe, instead of the parent being FeTe.
Therefore, the first conjecture considered is that there is a significant concentration of Fe vacancies in the epitaxial films. 
This would imply a lower concentration of Fe with respect to the stoichiometric condition.
On the other hand, if the vacancies are ordered, the superstructure should be detected by additional diffraction peaks and/or a magnetic ordering.
We couldn't find evidence of concentration being different from $1:1$, vacancy superstructure in XRD experiments, or a magnetic signal in magnetization measurements. Based on these results, we exclude the possibility of having a relevant concentration of ordered Fe vacancies. Nevertheless, the existence of a small concentration of vacancies not detectable within our resolution remains plausible.

To consider the conjecture that the deformation of the lattice leads to a significant change in the band structure, a crucial point is to determine which is the key physical parameter that quantifies the structural distortion. According to a recent report, there is a correlation between the semimetallic band gap and the in-plane strain $\epsilon=\sfrac{\Delta a}{a}$ in FeSe films ($-1.5\leq\epsilon\leq1.5$)\cite{Phan2017}. However, our epitaxial sample with $\epsilon=+0.64$ does not seem to follow this behaviour and is not semimetallic. Therefore, the in-plane strain by itself cannot explain a distortion of the electronic structure. Alternatively, the Poisson's ratio, $\nu=-\frac{\sfrac{\Delta c}{c}}{\sfrac{\Delta a}{a}}$, has been proposed to be the parameter that describes insulator-like FeSe epitaxial thin films \cite{HanzawaIEEE}.
To determine if there is a correlation between the insulator behavior and the structural distortion we compare the reported $\rho(T)$ curves of non-superconducting $\beta$-FeSe samples with the resistivity of our film.
First, we consider the reported curves from FeSe thin films fabricated with different growth methods (PLD \cite{Molatta2015}, MBE \cite{HanzawaPNAS}, Sputtering \cite{SchneiderPRL})
(Figure \ref{Transporte_FeSe_semi}). We found these curves to collapse onto a single curve when normalized by the value at $85\,$K, indicating a universal behavior. In the case of MBE's curve (Ref \cite{HanzawaPNAS}), $\nu\sim+75\,\%$. In contrast, our sample, shows a stronger semiconducting feature and has a higher Poisson's coefficient of $+162\,\%$. This positive correlation between the Poisson's coefficient and the semiconducting behavior may be indicative of a strong interplay between the distortion of the lattice and the band structure.
Additionally, we included the reported result for Fe$_4$Se$_5$ which has an even higher rise in resistance as the temperature decreases. This implies that the existence of vacancies produces an even more intense effect. In this case, with respect to the relaxed structure $\nu\sim-317\,\%$. This negative coefficient may be indicative of a non-trivial deformation. Therefore, the intermediate insulating behavior in our sample could be either related to structural distortion and/or a relevant vacancies concentration. 
Since we did not find a relevant concentration of Fe vacancies, the structural distortion may play the key role.

Regarding the conduction mechanism, there is a linear dependence of $\ln R_0$ with $T^{-\frac{1}{4}}$ in a wide range of temperature, being the exception the break at $\sim60\,$K for Fe$_4$Se$_5$ (see the inset in the Figure \ref{Transporte_FeSe_semi}).
We emphasize that the $T^{-\frac{1}{4}}$ dependence is more appropriate than other types of exponents.
The mathematical relation $\ln R\propto T^{-\frac{1}{4}}$ is widely used to describe the conductivity in strongly disordered systems with localized states (VRH model). Therefore, the dependence found may suggest a localization effect. The key question is which type of disorder originate the localization.
Recently, the formation of potential barriers in the conduction band has been suggested as the origin of the insulator-like behavior
\cite{HanzawaPRB}. In this scenario, despite a metallic electronic structure observed by ARPES, the carriers cannot move freely due to the influence of the potential barriers.

\section{Conclusions}

In summary, we have shown that the macroscopic electronic behavior of sputtered FeSe thin films is strongly susceptible to microscopic and mesoscopic characteristics as lattice distortion and/or grain morphology.
In the limit of textured thick films, we found optimal superconducting properties ($T_{\rm c}\sim12\,$K) at $T_{\rm s}= 530\,^\circ$C with a thickness around $400\,$nm. Characteristics as an enhanced $T_{\rm c}$, a relatively high $H_{\rm C2}$ and a low anisotropy are promising features. The enhancement of the critical temperature could be related to a nanoscale intergrowth of $\beta$-FeSe and $\gamma$-Fe$_7$Se$_8$ and/or stress associated with the coalescence of grains.
These samples allowed us to determine the sensitivity of some relevant physical properties to the presence of disorder. 
Properties like the coherence length and features like the change in the sign of the Hall coefficient concomitant with a positive magnetoresistance are robust to the amount of disorder present in our samples. Strikingly, the Hall effect is linear up to 16T, in contrast to the nonlinear behavior observed below T* in high-quality single crystals. 

For epitaxial stressed thin films, the characteristic semimetallic behavior disappears giving rise to an insulating one. We found that the structural distortion, described by the Poisson's coefficient, may play the key role instead of stoichiometric changes like ordered Fe vacancies. On the other hand, for a less distorted lattice restricted to independent grains, there is also a non-SC behavior. When these grains coalesce, due to thickness increase, superconductivity appears with higher $T_{\rm c}$ than that for bulk samples. This implies that the superconductor-insulator transition (SIT) induced by reducing the thickness can be understood taking in account the non-trivial evolution of the structural properties and morphology, which can be associated with the strained initial growth of the sputtered $\beta$-FeSe films.

\begin{acknowledgements}
We thank P. Troy\'on and M. Corte at CM-GIA-GAATEN-CNEA
for SEM/EDX characterization.
We also want to thank
S. Suarez and P. P\'{e}rez for the RBS measurements. Furthermore, we are grateful to 
M. Sirena and L. Avil\'{e}s for their help with the AFM characterization.
Work partially supported by Conicet PIP 2014-0164, ANPCyT PICT 2014-1265 and Sectyp UNCuyo 06/C441 and 06/C504.

\end{acknowledgements}

\nocite{*}
\bibliography{new}

\end{document}